\documentclass{iopjournal}
\usepackage{epsfig}
\usepackage{amssymb}
\usepackage{times}
\usepackage{amssymb}
\usepackage{amsmath}
\usepackage{cleveref}
\usepackage{mathrsfs}
\usepackage{graphicx}
\usepackage{epsfig}
\usepackage{dcolumn}
\usepackage{color}
\usepackage{bm}
\usepackage{graphics,psfrag}
\usepackage{graphicx,psfrag}
\usepackage{braket}
\usepackage{soul}
\newcommand{\be}{\begin{equation}}
\newcommand{\ee}{\end{equation}}
\newcommand{\bea}{\begin{eqnarray}}
\newcommand{\eea}{\end{eqnarray}}
\newcommand{\ba}[1]{\begin{array}{#1}}
\newcommand{\ea}{\end{array}}
\usepackage{graphicx,amsmath,amssymb,xcolor,hyperref}
\begin{document}
 \title{Ion-atom two-qubit quantum gate based on phonon blockade}
 \author{Subhra Mudli and Bimalendu Deb$^{*}$}\\
 \email{msbd@iacs.res.in}\\
 \affil{School of Physical Sciences, Indian Association for the Cultivation of Science, Jadavpur, Kolkata 700032, India.}\\
\begin{abstract}
 We theoretically demonstrate the universal two-qubit CNOT gate between an ionic and an atomic qubit relying on Rydberg excitation of the atom and the resulting phonon blockade in the motional states of the harmonically trapped ion. The phonon blockade arises due to strong ion-atom interaction when the atom is excited to a Rydberg state. For realistic parameters, the gate fidelity is found to be about $90\%$. In a previous paper [S. Mudli {\it et al.} Phys. Rev. A 110, 062618 (2024)], it was shown that a trapped ion can mediate interaction between two largely separated Rydberg atoms, and this mediated interaction can be leveraged to perform a universal two-qubit gate operation between neutral atom qubits in optical tweezers. These demonstrations suggest that an ion-atom hybrid system can serve as a resourceful platform or module for quantum computing and quantum networking as it can utilize the best features of charged as well as neutral atom qubits. Finally, we discuss how to achieve higher gate fidelity by extending our proposed protocol and operating in a different parameter regime.
\end{abstract}

\section{Introduction}

Among quantum computing platforms, superconducting circuit QED \cite{Frunzio:IEEETOAS:2005, Blais:PRA:2004, Blais:PRA:2007, Blais:RMP:2021},
trapped ions \cite{Monroe:SA:2008, Blatt:Nature:2008,  Blatt:NatPhys:2012, Bruzewicz:APR:2019} and neutral atoms \cite{Negretti:QIP:2011, Weiss:PT:2017, Henriet:Quantum:2020, Graham:Nature:2022, Evered:Nature:2023, Bluvstein:Nature:2024} are the forerunners as they have long qubit coherence times and allow precise control over their quantum states. Trapped ions are employed to create high-fidelity laser-driven quantum gates mediated by their quantized motional modes. Advancements in trapped-ion technology have significantly improved both state preparation and quantum gate performance. A recent theoretical proposal for Doppler cooling of trapped ions with $\Lambda$-type three-level configuration \cite{Yan:PRA:2019} and the experimental demonstration of nonadiabatic noncyclic fast and robust single-qubit geometric gate operation \cite{Zhang:PRL:2021} highlight the level of precision that can be achieved in trapped-ion systems. Neutral atoms when excited to Rydberg states  undergo strong long-range dipole-dipole (dd) or van der Waal (vdW) interactions that can be engineered on a fast time scale using well-designed laser pulses \cite{Saffman:RMP:2010, Isenhower:PRL:2010, Beguin:PRL:2013, Browaeys:NatPhys:2020, Wang:PRX:2022, Zhang:PRX:2022,  Rej:Arxiv:2025, Rej:NJP:2026, Rej:Arxiv:2026}. Although, recent experimental works on merging cold atoms with ions in a Paul trap \cite{Ewald:PRL:2019, Secker:PRL:2017, Tomza:RMP:2019} have demonstrated intriguing features of ion-atom cold collisions \cite{Rakshit:PRA:2011}, creating an ion-atom hybrid architecture for quantum computation \cite{Mudli:PRA:2024} by combining atom optical tweezers with an ion-trap remains a challenging problem. Any hybrid quantum systems that combine such distinct physical platforms provide new exploratory frameworks for emerging quantum technology \cite{Buluta:RPP:2011, Xiang:RMP:2013}. A recent experiment has successfully inserted an empty optical tweezer inside an ion trap \cite{Mazzanti:PRA:2024}. Thus, one expects that in near future it will be possible to make a hybrid array of ions in a Paul trap and atoms in optical tweezers.

Our ion-atom two-qubit gate protocol is schematically shown in Fig \ref{fig 1}. In this proposed model, a single neutral atom is trapped in an optical tweezer positioned near a trapped ion in a Paul trap. We do not want the optical tweezer to be too close to the ionic equilibrium position to avoid direct atom-ion collision. The atomic qubit is encoded in two hyperfine ground states while an auxiliary Rydberg state of the atom is used to mediate conditional dynamics. The ionic qubit is encoded in two internal states and coupled to its quantized motional states via laser-driven sideband transitions. When this atom is excited to a Rydberg state, it results in a long-range ion-atom polarization interaction that modifies the trapping potential of the ion. This interaction leads to shifts in both the equilibrium position and the vibrational frequency of the ion's harmonic confinement. Since trapped ion quantum logic relies critically on the resonance condition between internal electronic transitions and quantized motional sidebands, such interaction-induced shifts can be used as a resource for hybrid ion-atom quantum computation. The shifts effectively detune the red-sideband transition suppressing phonon exchange when the detuning exceeds the ion-phonon coupling. We call this suppression effect as phonon blockade. By exploiting this conditional phonon blockade, we design a controlled-NOT (CNOT) gate in which the atomic qubit acts as the control and ionic qubit as the target. The gate is implemented through a sequence of laser pulses. The atomic qubit is first conditionally excited to a Rydberg state followed by a sideband pulse on the ion. The transition between the ionic qubit states is either allowed or blocked depending on the qubit state of the atom. Note that, usually phonon blockade refers to a nonlinear effect in nano or micro-mechanical oscillator systems by which the emergence of a second phonon is blocked by the first phonon. This is analogous to photon blockade in quantum optics. Here our phonon blockade has a different meaning - it is the large shift in the phonon frequency that blocks the excitation of a phonon in the ionic vibrational quantum states by a two-photon Raman process.

We develop a detailed theoretical model \cite{Mudli:PRA:2024} of the atom-ion system including ion-phonon coupling, laser driven internal dynamics, and the Rydberg mediated atom-ion interaction. By deriving an effective Hamiltonian in Lamb-Dicke regime, we identify the parameter conditions under which phonon blockade occurs and unwanted transitions are suppressed. Our numerical analysis  using $^{87}$Rb$-^{9}$Be$^{+}$ hybrid system with realistic parameters show that an ion-atom two-qubit gate with about 90$\%$ fidelity can be realized within a time scale that is much shorter than the lifetime of the Rydberg state. This work shows phonon blockade as a  resource for quantum logic in a hybrid atom-ion platform,  highlighting the importance  of atomic Rydberg state-mediated ion-atom interactions,  apart from ion-mediated atom-atom interactions \cite{Mudli:PRA:2024} for quantum computing.

The paper is organized in the following way. In Sec \ref{Sec:2} we present the model and formulate the problem. In Sec \ref{Sec:3} we show numerical results and interpret them. We conclude in Sec \ref{Sec:4}. Finally, we discuss the possibility of increasing the gate fidelity by using stronger Rabi coupling to multiple Rydberg states.

\section{ The model and formulation of the problem}\label{Sec:2}
\begin{figure}
\hspace{-0.97in}
  \begin{center}
  \includegraphics[height=2.4in,width=3.5in]{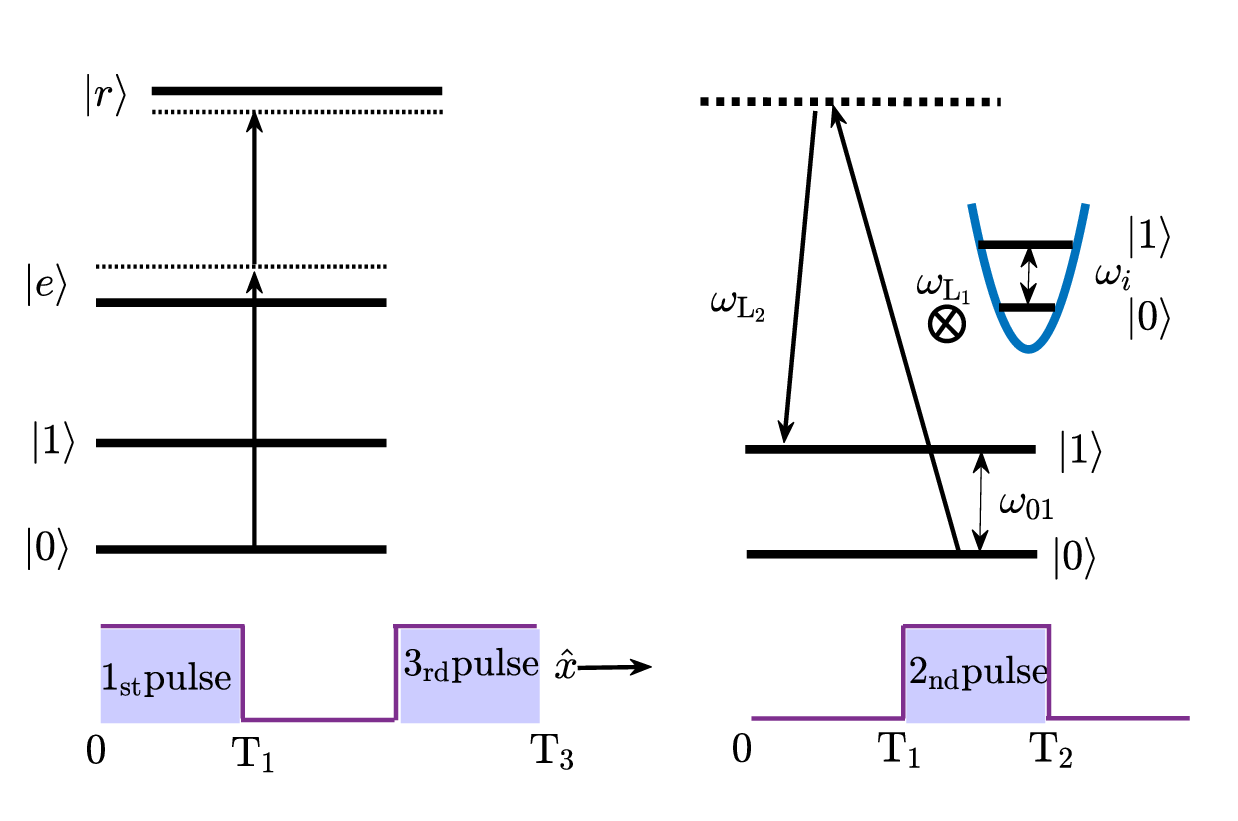}\\
  \end{center}
  \caption{A schematic of control-NOT gate using phonon blockade: A single atomic qubit (the left level diagram) is placed at a small separation from an ionic qubit (the right level diagram). The atom is assumed to be trapped in an optical tweezer which is positioned near  the ion  in a Paul trap. Two hyperfine levels in the ground state manifold  constitute the two atomic qubit states $|0\rangle$ and $|1\rangle$. The ionic qubit may be composed of two ground-state hyperfine levels or one ground-state hyperfine level and one metastable excited level. The  tweezer is considered to be collinear with the $x$-axis of the Paul trap. Here controlled NOT gate is realized via phonon blockade through controllable ion-atom interaction. When the atomic qubit state $|0\rangle$ is coupled to the Rydberg state $|r\rangle$ by a two-photon process via the intermediate state $\mid e \rangle$, the ion-atom interaction is enhanced by many orders of magnitude. Ion's qubit states are coupled to its phonon states by a two-photon laser coupling, with $\omega_{L_1}$ and $\omega_{L_2}$ being the frequencies of the two lasers. First, a two-photon $\pi$ pulse acts on the control (atom), then a $\pi$ rotation is given to the target (ion), lastly another $\pi$ pulse is applied on the control. By this pulse sequence,  a controlled NOT gate is realized (see text).}
  \label{fig 1}
\end{figure}

A schematic diagram  of our proposed model is shown in Fig.\ref{fig 1}. An atom is trapped in an optical tweezer whose center is placed at a distance $x_{0}$ from the center of a Paul trap containing a single ion. The electric potential minimum of the ion-trap or the equilibrium position of the ion is taken as the origin.

To begin with, let us consider that the internal (electronic) states of the atom is $|0 \rangle \equiv (1 \hspace{0.2cm} 0 )^{T}$, $|1 \rangle \equiv (0 \hspace{0.2cm} 1 )^{T}$ and $|r\rangle$; where $|0 \rangle$ and $|1 \rangle$ denote two ground-state hyperfine levels and $|r \rangle$ a Rydberg state.
The atomic qubit is composed of  $|0 \rangle$ and  $|1 \rangle$. $| r \rangle$ will be employed as an auxiliary state for performing two-qubit quantum gate operations. The ionic qubit consists of $|0 \rangle$ and  $|1 \rangle$, which may be two grounded hyperfine levels or a ground hyperfine state and a metastable excited state.

For simplicity of calculations, we consider an effective one-dimensional (1D) model system along the $x$-direction, assuming the harmonic trapping frequencies of both the ion trap and the optical tweezer along $y$ and $z$ directions are much higher than that along $x$ direction and both  the ion  and the atom are cooled to the motional ground states of harmonic oscillations along $y$ and $z$ direction.  The ion-atom interaction Hamiltonian is $\hat{V}_{\rm{ia}}(x_a, x_i) = \hat{V}_{\rm{ia}}(|x_a-x_i|)$ where
\begin{equation}\label{eq:1}
\hat{V}_{\rm{ia}} =  V  |r\rangle \langle  r| \otimes  | i \rangle \langle  i |
\end{equation}
Here $V = \frac{C_{4}}{|x_a - x_i|^4}$ with $C_{4}$ being the long-range coefficient of interaction between the ion and the atom in the Rydberg state. Since $|x_a| >\!> |x_i|$, we have
 \bea
 V \simeq \frac{C_4}{x_a^4} \left [ 1 + \frac{4 x_i}{x_a}  + \frac{4 x_{i}^{2}}{x_{a}^{2}}\right ] = V^{(0)} + \hat{U}_{1} + \hat{U}_{2}
 \label{eq:2}
 \eea
 where $V^{(0)} = C_4/x_a^4$ and
\bea \label{eq:3}
\hat{U}_1 &=& U_1^{(0)} \frac{1}{\sqrt{2}} \left [ a_i e^{- i \omega_i t} + a_i^{\dagger} e^{i \omega_i t} \right ]
\eea
\bea \label{eq:4}
\hat{U}_2 &=& U_2^{(0)} \left [ a_i^{2} e^{- 2i \omega_i t} + a_i^{\dagger 2} e^{2i \omega_i t}  + a_{i}a_{i}^{\dagger} + a_{i}^{\dagger}a_{i} \right]
\eea
with $U_{1}^{(0)} = V^{(0)}\beta$ and $U_{2}^{(0)} = V^{(0)}\beta^{2}/8$ where $\beta = 4 \sqrt{2} \lambda_i/x_{a}$ is the ratio between the width $\lambda_i$  of the motional ground-state probability function  of the ion and the position $x_{a}$ of the atom. Here we have written  the position of the ion $x_{i}$ in the quantized form $x_{i} =  \lambda_i \left [ a_i e^{- i \omega_i t} + a_i^{\dagger} e^{i \omega_i t} \right ] $,   where $\omega_i$ is the harmonic trapping frequency, $\hat{a}_i^{\dagger} (\hat{a}_i )$ represents the creation (annihilation) operator of the 1D harmonic motional quanta (phonon), and $\lambda_i = \sqrt{\frac{\hbar }{ m_i \omega_i}}$ with $m_i$ being the mass of the ion. Note that we work in a frame where the ionic phonon operators rotate with unperturbed harmonic frequency $\omega_{i}$ of the ions motional oscillations.  The atom-ion interaction of Eq. (\ref{eq:2}) changes the potential minimum and phonon frequency of the ion trap. The modified frequency and equilibrium position of the ion trap is given by $\bar{\omega}_{i}^{2} = \omega_{i}^{2}-\frac{8C_{4}}{m_{i}x_{a}^{6}}$ and $\bar{x} = x_{i} - \frac{4C_{4}}{m_{i}\bar{\omega}_{i}^{2}x_{a}^{5}}$,  respectively.

\begin{figure}
\hspace{-0.97in}
  \begin{center}
    \includegraphics[height=2.2in,width=3in]{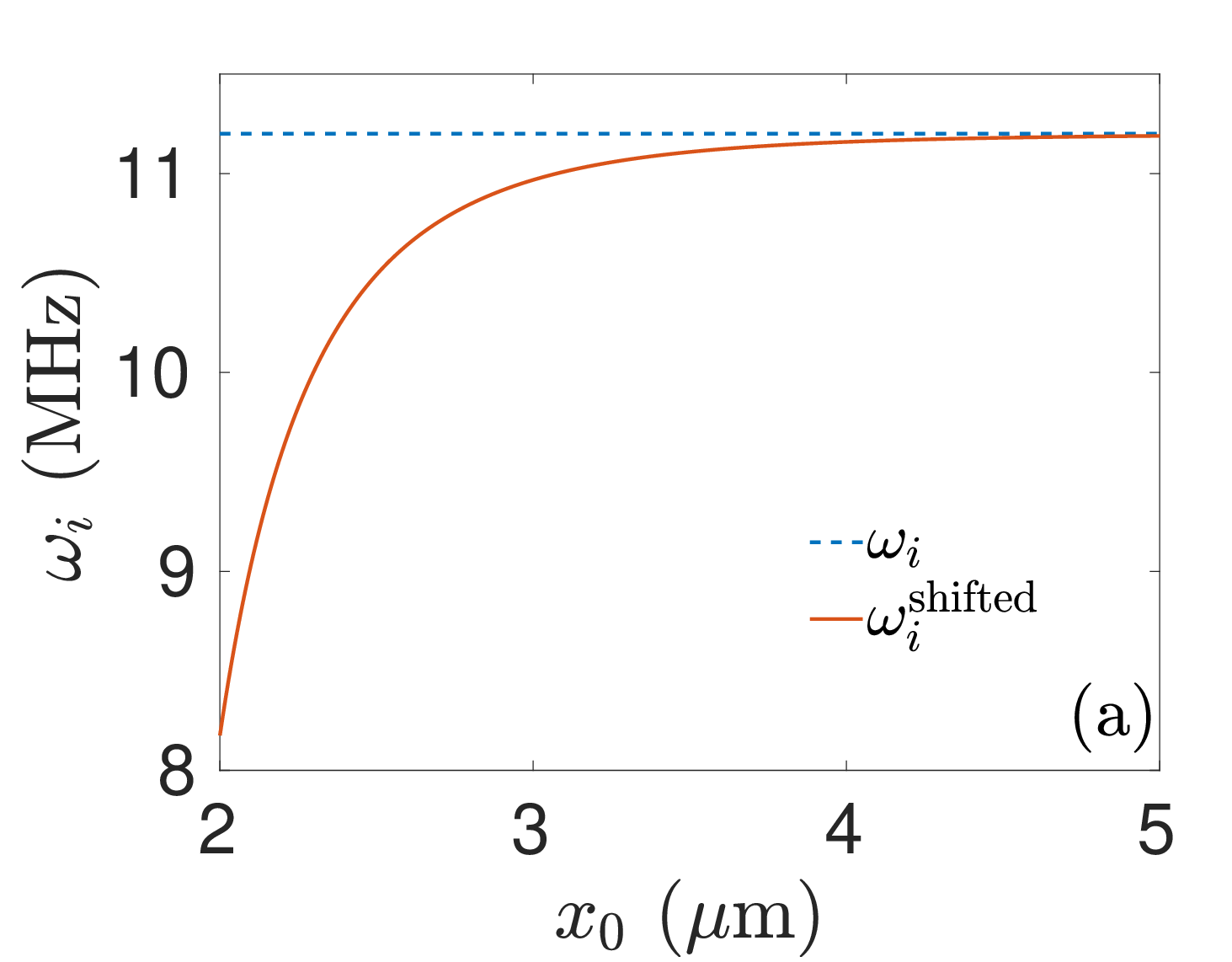}
    \includegraphics[height=2.2in,width=3in]{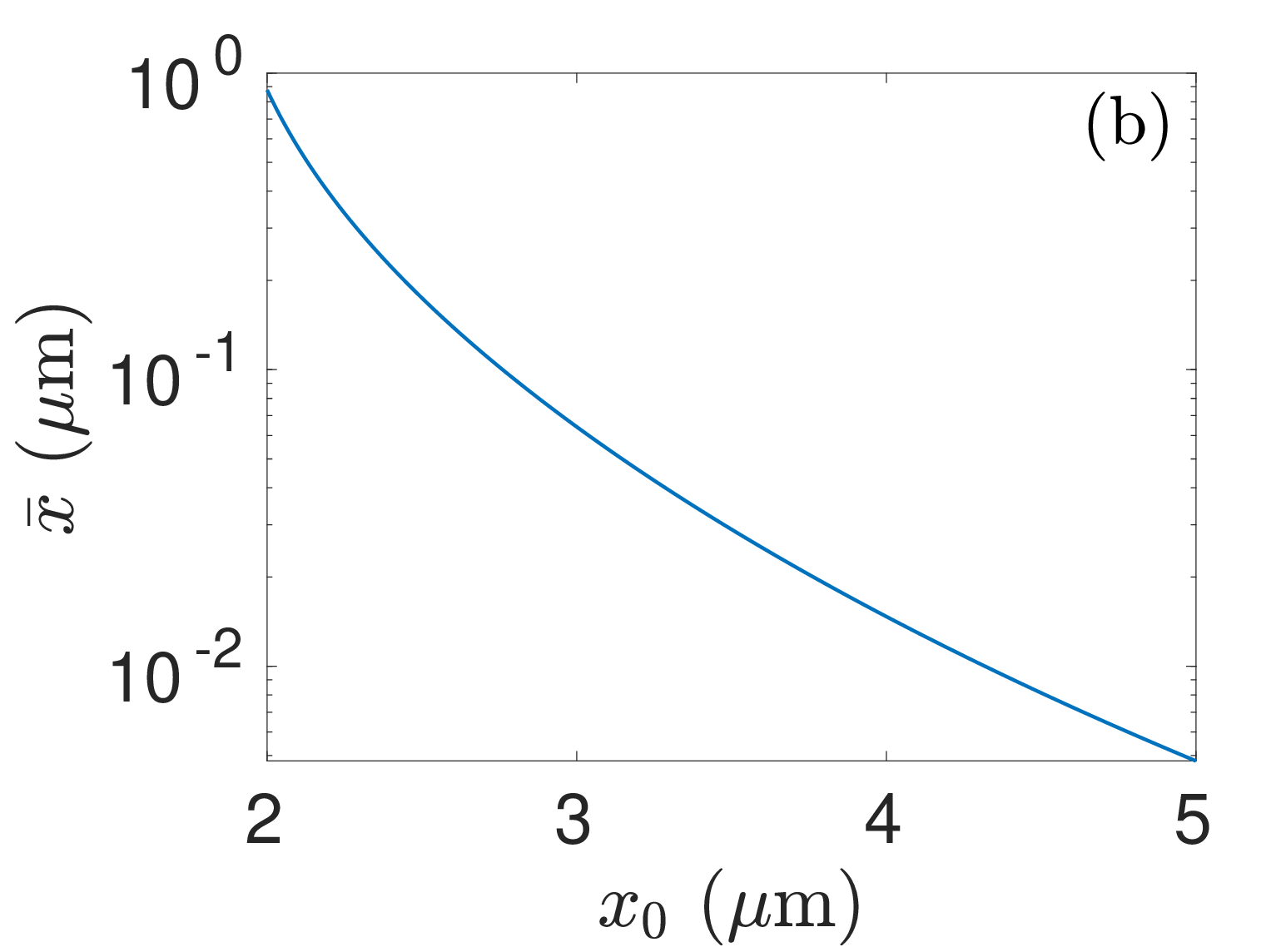}
  \end{center}
  \caption{The shifted  phonon frequency (in MHz) (a) and  the shifted equilibrium position (in $\mu$m) (b) are plotted as a function of ion-atom distance $x_0$ (in $\mu$m) for $^{87}$Rb+$^{9}$Be$^{+}$ system, when the atom is excited to the Rydberg state with principal quantum number $n=90$.  The ion's unperturbed trapping frequency (dashed line) is $2\pi \times 11.2$ MHz.}
  \label{fig 2}
\end{figure}

\subsection{Phonon blockade}

In an ion-atom hybrid  quantum system that we consider, phonon blockade manifests through the ion-phonon interaction mediated by atom-ion coupling. When the effective detuning for the ionic motional sideband transitions exceeds the phonon-ionic qubit coupling strength, the transitions between phonon number states becomes off-resonant. Consequently, the emission or absorption of a phonon is suppressed. As a result, a blockade in the transition between the two ionic qubit states will occur.

To realize the phonon blockade and quantum gates, consider that the qubit state $|0 \rangle $ of the atom is coupled to the Rydberg state $|r \rangle$ by a two-photon laser pulse. The Hamiltonian describing the system is given by $\hat{H} = \hat{H}_{i} + \hat{H}_{a} + \hat{U}_{\rm{ia}} $ where
\bea
\hat{H}_{i} = \frac{\hbar\omega_{\rm 01}\sigma_{z}}{2} + \hat{H}^{(m)} + \hat{H}_{L}^{(i)},
\label{eq:5}
\eea
\bea
\hat{H}^{(m)} = -\frac{\hbar^{2}}{2m_{i}}\frac{d^{2}}{dx_{i}^{2}} + \frac{1}{2}m_{i}\omega_{i}^{2}x_{i}^{2}
\label{eq:6}
\eea
describes COM motional dynamics of the ion and $\omega_{01}$ is the ionic qubit transition frequency. Here
\bea
\hat{H}_{L}^{(i)}=\frac{1}{2}\hbar \Omega_{i} \left(\sigma_{+} + \sigma_{-}\right)\times\left[e^{i(kx_{i}-\omega t + \phi)}+e^{- i (kx_{i}-\omega t + \phi)}\right]
\label{eq:7}
\eea
is the interaction of the ionic qubit with the laser, with $\omega = \omega_{L_1} - \omega_{L_2}$ being the difference between the frequencies of the two lasers that couple the two ionic qubit states.

We take the two hyperfine states $|0\rangle$ and $|1\rangle$ to be the qubit states of the ion. So we can represent $\sigma_{+} \equiv |1\rangle\langle 0|$ and $\sigma_{-} \equiv |0\rangle\langle 1|$. $\Omega_{i}$ is the Rabi frequency for transition $|0 \rangle \leftrightarrow |1 \rangle $. Here the atomic Hamiltonian is
\bea
\hat{H}_{a} = \hat{H}_{0}^{(a)} + \hat{H}_{L}^{(a)}
\eea
where
\bea
\hat{H}_{0}^{(a)} = - \hbar \delta_{a} |r \rangle \langle r | \otimes \mathbb{I}_{i}\nonumber
\eea
\bea
\hat{H}_{L}^{(a)} &=& \frac{1}{2}\hbar \left [ \Omega_a |r \rangle \langle 0 | + {\rm H.c.}  \right ] \otimes \mathbb{I}_{i}
\label{eq:8}
\eea
where $\mathbb{I}_{i} =  |0\rangle\langle 0| +  |1\rangle\langle 1| $. Here $\hat{U}_{ia} = \hat{U}_{1} + \hat{U}_{2}$ is the ion-atom interaction Hamiltonian. $\hat{H}_{L}^{(a)}$ is the Hamiltonian that describes interaction of the atomic qubit with lasers and $\delta_{a} = \delta_{r}-\frac{V_{0}}{\hbar}$, $\delta_{r} = \omega_{{\rm sum}} - \omega_{ra}$ with $\omega_{{\rm sum }}$ being the sum of the frequencies  of the two lasers that couple the state $|0 \rangle$ with  $|r \rangle$ and $\omega_{ra}$ is the atomic frequency for $\mid 0 \rangle \leftrightarrow \mid r \rangle$ transition. Here $\Omega_a$ is the two-photon Rabi frequency for transition $|0 \rangle \leftrightarrow |r \rangle $. $\hat{U}_{ia}$ is the ion-atom interaction Hamiltonian.

The unitary operator for the atomic part of the Hamiltonian is
\bea
U_{a}^{(1)}(t)&=&e^{-i\int_{0}^{t}\left(\hat{H}_{0}^{(a)}+\hat{H}_{L}^{(a)}+\hat{U}_{ia}(t')\right)dt'/\hbar}\nonumber\\
&=& e^{-i(\eta(t)\sigma_{0}^{a}+\Omega_{a}t\sigma_{x}^{a}+\eta(t)\sigma_{z}^{a})/2}\nonumber\\
&=& e^{-i\eta(t)\sigma_{0}^{a}}\left[\cos{\frac{\hat{\theta}}{2}}\mathbb{I} -i\frac{\hat{\eta}(t)}{|\hat{\theta}|}\sin{\frac{\hat{\theta}}{2}}\sigma_{z}^{a} -i\frac{\Omega_{a}t}{|\hat{\theta}|}\sin{\frac{\hat{\theta}}{2}}\sigma_{x}^{a}\right]
\eea
where $\hat{\theta}=\sqrt{|\hat{\eta}|^{2}+|\Omega_{a}|^{2}t^{2}}$ and $\hat{\eta}(t)=\int_{0}^{t}(\delta_{a}-\hat{U}_{1}(t')-\hat{U}_{2}(t'))dt'$. Let the initial state be $|000\rangle$ where the atomic, ionic and phononic states are written, respectively. Then,  on application of the two-photon laser pulse on the atom for a duration of $t$, the atom-ion-phonon joint state will be evolved to
\bea
|\psi(t)\rangle &=& e^{-i\eta(t)\sigma_{0}^{a}}\left[\cos{\frac{\hat{\theta}}{2}}|000\rangle -i\frac{\hat{\eta}(t)}{|\hat{\theta}|}\sin{\frac{\hat{\theta}}{2}}|000\rangle -i\frac{\Omega_{a}t}{|\hat{\theta}|}\sin{\frac{\hat{\theta}}{2}}|r00\rangle\right]
\label{eq:9}
\eea
If we want to make the transition $|0\rangle \rightarrow |r\rangle$ we have to consider $\Omega_{a} \gg U_{1}^{(0)}$ as $\hat{\eta}_{max} \propto U_{1}^{(0)}$ such that $\frac{\Omega_{a}t}{|\hat{\theta}|} \approx \left(1-\frac{\hat{\eta}^{2}}{2\Omega^{2}t^{2}}\right) \approx 1$ and $\frac{\hat{\eta}}{|\hat{\theta}|} \approx \frac{\hat{\eta}}{\Omega_{a}t}\left(1-\frac{\hat{\eta}^2}{2\Omega_{a}^{2}t^{2}}\right) \ll 1$. Otherwise due to high ion-atom interaction potential higher phonon states would be coupled, so the probability for Rydberg excitation would be reduced.
 We transform the ionic Hamiltonian into interaction picture under rotating wave approximation \cite{Liebfried:RMP:2003} in Lamb-Dicke regime
\bea \label{eq:10}
H_{int}&=&\frac{1}{2}\hbar\Omega_{i}\sigma_{+}\left\{1+i\eta\left(ae^{-i\omega_{i}t}+a^{\dagger}e^{i\omega_{i}t}\right)\right\}e^{i(\phi-\delta_{i} t)} + H.c.
\eea
Lamb-Dicke parameter $\eta=kx_{0}$, $x_{0}=\sqrt{\hbar/2m_{i}\omega_{i}}$ is the length scale and $\omega_{i}$ is the trapping frequency of ion. Here $\delta_{i}=\omega-\omega_{01}$. Now if we take $\delta_{i}=-\omega_{i}$ the Hamiltonian is first red sideband resonant, so the Hamiltonian becomes
\bea
\hat{H}_{i} &=& \frac{1}{2}\hbar \Omega_{i}\eta \left [a_{i}\sigma_{+}e^{i\phi} + {\rm h.c.}  \right ]
\label{eq:11}
\eea
where $\phi$ is a constant phase. Now if,  the atom is in  state $|1 \rangle$,  the  red sideband resonance in ion  holds good, and  ion-phonon transition will occur as a result. Now, if the atom is in Rydberg state, the shifts in the ionic equilibrium position and phonon frequency are large because the ion-atom interaction term increases as $n^{7}$ where $n$ is the principal quantum number of Rydberg state. When the shifted phononic frequency is $\bar{\omega}_{i} = \omega_{i} - \Delta$, $\Delta$ is the frequency shift, the ionic Hamiltonian becomes
\bea
\hat{H}_{i} &=& \frac{1}{2}\hbar \Omega_{i}\eta \left [a_{i}\sigma_{+}e^{i\left(\phi+\Delta\right)t} + {\rm h.c.}  \right ]
\label{eq:12}
\eea
This Hamiltonian can be written in time-independent form using a frame transformation. Then the effective Hamiltonian is
\bea
\hat{H}_{i}^{eff} &=& \frac{1}{2}\hbar \Omega_{i}\eta\sqrt{n} \sigma_{x} - \frac{\hbar\Delta}{2} \sigma_{z}
\label{eq:13}
\eea
As we consider first red sideband coupling, we use only two ion-phonon states which are $|0, n\rangle$ and $|1, n-1\rangle$, where $n$ is the phonon number. The unitary operator for this effective Hamiltonian is
\bea
\hat{U}_{i} &=& \cos{\frac{\theta_{i}}{2}}\mathbb{I}+i\frac{\Delta }{\sqrt{\Omega_{n}^{2} +\Delta^{2} }}\sin{\frac{\theta_{i}}{2}}\sigma_{z}-i\frac{\Omega_{n}}{\sqrt{\Omega_{n}^{2} + \Delta^{2} }}\sin{\frac{\theta_{i}}{2}}\sigma_{x}
\label{eq:14}
\eea
where $\Omega_{n}=\Omega_{i}\eta\sqrt{n}$,  $\theta_{i}= t \sqrt{\Omega_{n}^{2} +\Delta^{2} }$. Equation (\ref{eq:14}) shows  that if $|\Delta| >> \Omega_{n}$ transition between the levels is highly suppressed. In this limit, if we make $\theta_{i}=\pi$, then we have $\hat{U}_{i}\simeq i \sigma_{z}=e^{i\pi/2}\sigma_{z}$.

\begin{figure}
\hspace{-0.97in}
 \begin{center}
    \includegraphics[height=2.2in,width=5.5in]{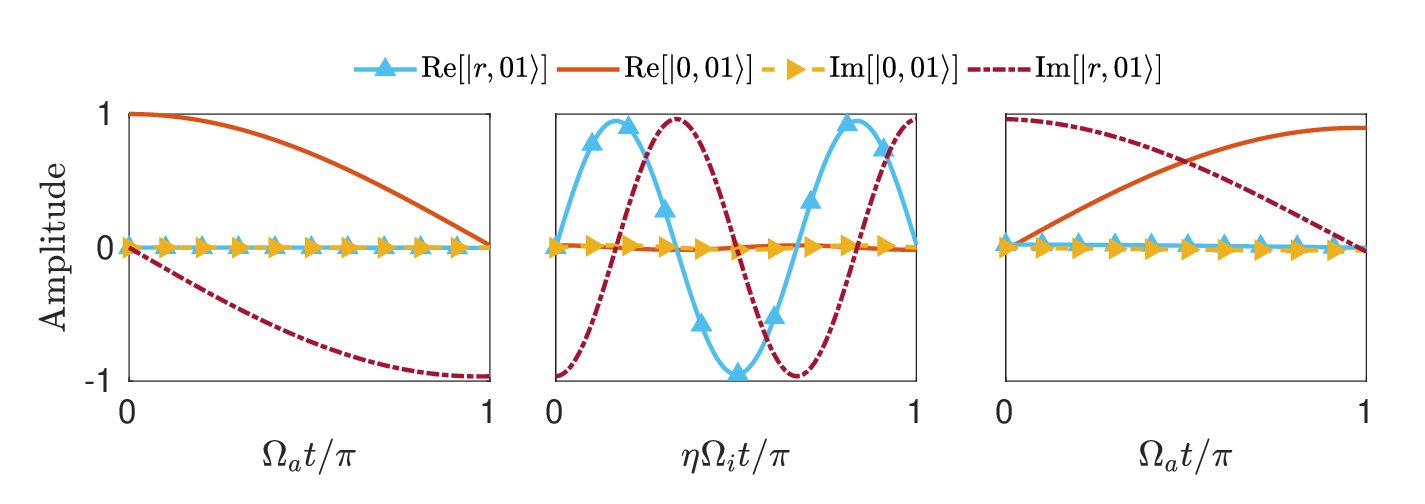}\\
    \includegraphics[height=2.2in,width=5.5in]{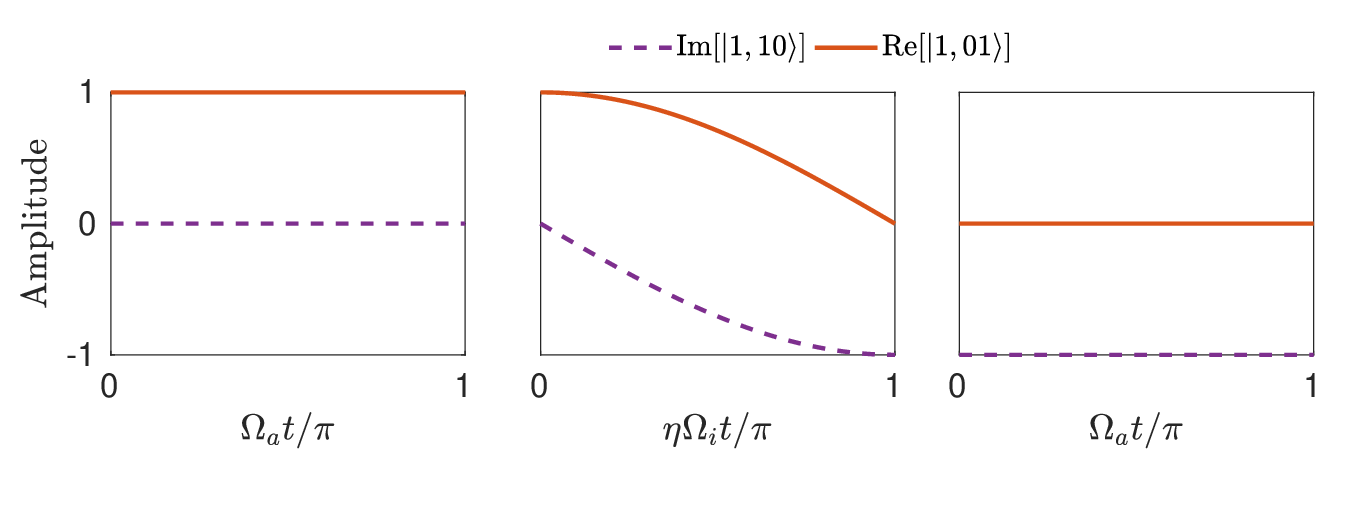}
  \end{center}\vspace{-0.2in}
  \caption{ The  real and imaginary parts of the amplitude of different states $|\rm{a}, \rm{i ph}\rangle$ are plotted as a function of dimensionless time. First and third columns show the evolution of states during the first and third laser pulses applied on the control (atom) qubit while the second column displays the evolution of state during the second pulse applied on the target (ion) qubit. Note that during the first and third pulses, the unit of time is $\pi/\Omega_{a}$ while that during the second pulse is $\pi/(\eta\Omega_{i})$. The first row of three sub-figures show that when the control (atom) is initially in $|0\rangle$ and the target (ion) is in $|0\rangle$ with $1$ phonon, after first $\pi$-pulse, the atom is excited to Rydberg state with a phase being almost equal to $\pi$. Now,  due to Rydberg excitation and the resulting phonon blockade, transitions between the phononic states are blocked during the action of the $\pi$-pulse on the target. Thereafter, the evolved states return to initial states upon application of another $\pi$- pulse on the control. The second row of three sub-figures show the evolution of the states when  the control  is initially in $|1\rangle$ and the target  is in $|0\rangle$ with $1$ phonon, here as the atom is not excited to the Rydberg state,  transitions between the ionic states occur resonantly.}
  \label{fig 3}
\end{figure}

\subsection{Ion-atom two-qubit gate protocol based on phonon blockade}

We use this phonon blockade to realize CNOT gate operation. CNOT gate implements the transformation
\begin{eqnarray}
 \left\{ |00 \rangle, |01 \rangle, |10 \rangle, |11 \rangle \right\} \rightarrow \left\{|00 \rangle, |01 \rangle, |11\rangle, |10\rangle \right\}
 \label{eq:15}
\end{eqnarray}
where the atom acts as control and the ion as target. Our proposed gate protocol
is composed of three consecutive pulses, first pulse is to excite control qubit to a Rydberg state depending on the initial atomic qubit state, second pulse is to excite target qubit conditioned on the control state and third pulse is to deexcite the control qubit if it is in Rydberg state. Here gate protocol is discussed step by step: \\
{\bf Step I}: the application of the first $\pi$ pulse on the control qubit is governed by the Hamiltonian
$\hat{H}_{1}=\hat{H}_{a}$. For different initial state preparations, one should ideally observe the following evolution after the first $\pi$ pulse:
\begin{eqnarray}
 |0,01 \rangle \rightarrow -i|r,01 \rangle, |0,10 \rangle \rightarrow -i|r,10 \rangle, |1,01 \rangle \rightarrow |1,01 \rangle, |1,10 \rangle \rightarrow |1,10 \rangle
 \label{eq:16}
\end{eqnarray}
implying that if the atom is initially in $|0\rangle$ state irrespective of the state of the ion, the first $\pi$ pulse will excite the atom to the Rydberg state with a phase $-\pi/2$ while nothing will happen if the atom is initially prepared in $|1\rangle$. We define $|\rm{a},\rm{i ph}\rangle \equiv |i\rangle\otimes|a\rangle\otimes|\rm{ph}\rangle$ where $|a\rangle, |i\rangle \rm{and} |\rm{ph}\rangle$ define atom, ion and phonon basis, respectively.\\
{\bf Step II}: The atomic state does not change during second pulse but it can aquire a phase. The Hamiltonian for second pulse is $\hat{H}_{2}=\hat{H}_{i}$. This pulse is applied for a duration of $\pi/(\eta\Omega_{i})$. After which the states $-i|r,01\rangle$, $-i|r,10\rangle$, $|1,01\rangle$ and $|1,10\rangle$ should ideally evolve to $-ie^{i\phi}|r,01\rangle$, $-ie^{i\phi}|r,10\rangle$, $-i|1,10\rangle$ and $-i|1,01\rangle$, respectively where the phase $\phi$ depends on the shift $\Delta$ and $\Omega_{i}$. We  choose the distance and $\Omega_{i}$ such that $\phi=\pi$. \\
{\bf Step III}: The Rydberg atom is deexcited to the ground state $|0\rangle$ by the application of a pulse for duration $\pi/\Omega_{a}$. So the final states should become $|0,01\rangle$, $|0,10\rangle$, $-i|1,10\rangle$ and $-i|1,01\rangle$, for a particular ion-atom distance and particular $\Omega$ and $\phi = \pi$. Now if a single qubit phase gate, $\hat{S}= \begin{pmatrix}
                                                                                                                                                                                                                                                        1 & 0\\
                                                                                                                                                                                                                                                        0 & i\\
                                                                                                                                                                                                                    \end{pmatrix}
$ is applied on control we get the final states $|0,01\rangle$, $|0,10\rangle$, $|1,10\rangle$ and $|1,01\rangle$. Thus one can realize an ion-atom CNOT gate. Here the $\hat{S}$ gate changes the phase of the control qubit state $|1\rangle$ by $\pi/2$ while the state $|0\rangle$ of the qubit remain unaffected. This $\hat{S}$ gate is necessary to eliminate the extra phase $-\pi/2$ acquired by the both states of target qubit when the control qubit is in $|1\rangle$ during the second $\pi$ pulse.

\section{Results and discussions}\label{Sec:3}

For numerical illustration, we consider  {$^{87}$Rb+$^{9}$Be$^{+}$} system. The atomic qubit is encoded in two hyperfine ground states,  $|0\rangle \equiv |n=5, L=0, F= 1\rangle$ and $|1\rangle \equiv |n=5, L=0, F= 2\rangle$, while an auxiliary Rydberg state $|r \rangle \equiv |n=90, L=0, F=2 \rangle $ is used to mediate conditional dynamics. The chosen Rydberg level has a lifetime approximately 100 $\mu$s, which provides sufficiently long coherence window for gate operations with nanosecond to microsecond scale pulses \cite{Wang:PRX:2022}. For this highly excited  Rydberg state, the ion-atom interaction is significantly enhanced. The corresponding long-range interaction coefficient $C_{4}$ is $5.07 \times 10^{10}C_{4}^{0}$ \cite{Kamenski:JPB:2014} where $C_{4}^{0} = -160 a.u.$. This large enhancement originates from the strong polarizability of the Rydberg atom, scaling as $n^{7}$, and ensures that even at separations of several microns, the ion's motion is strongly influenced when the atom is excited to $|r\rangle$. For the ion, the qubit states are two hyperfine ground states, $|0\rangle \equiv |n=2, F=2\rangle \rm{and} |1\rangle \equiv |n=2, F=1\rangle$ \cite{Wineland:PRL:1996}.

\begin{figure}
\hspace{-0.97in}
 \begin{center}
    \includegraphics[height=2.5in,width=3in]{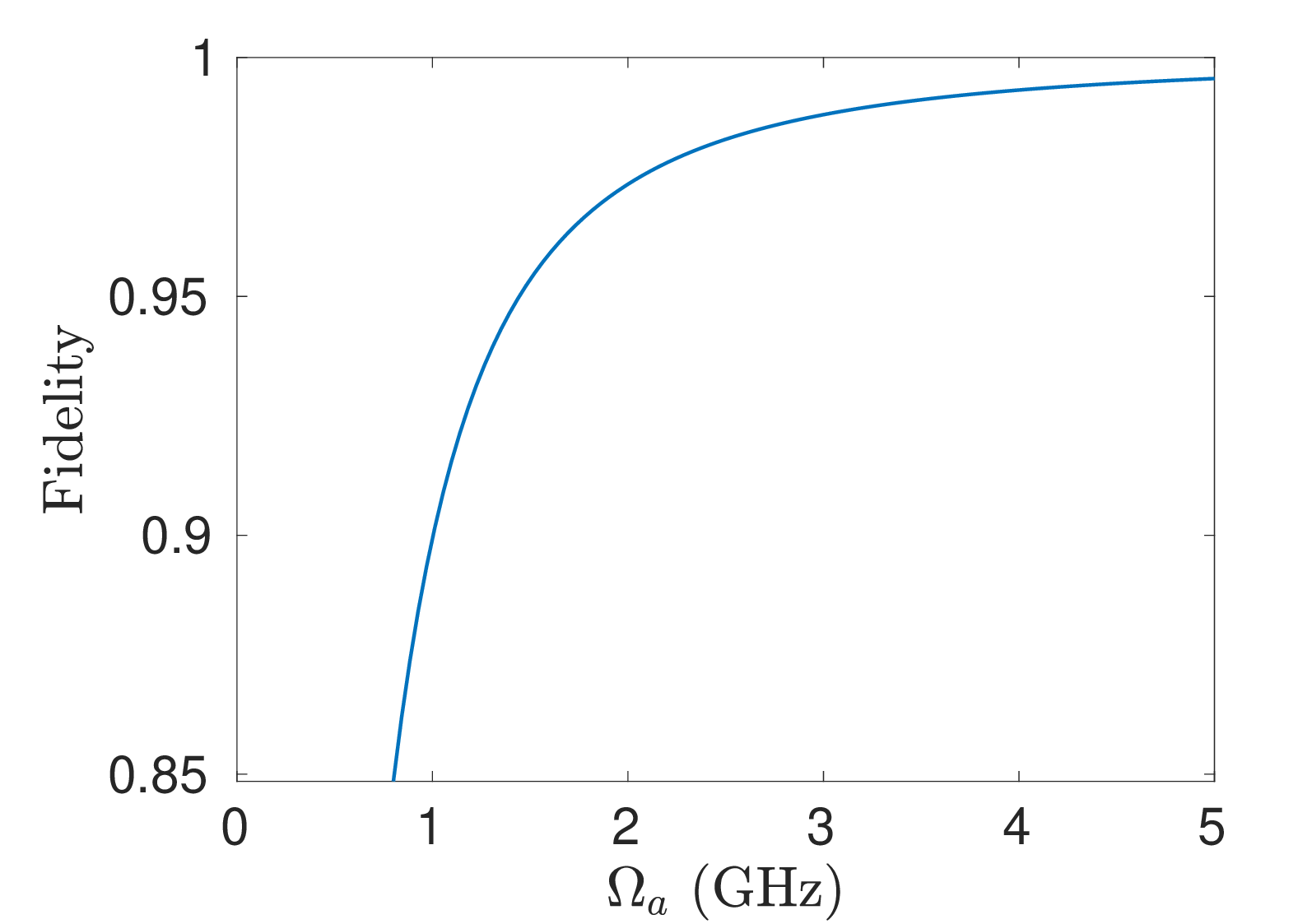}
  \end{center}\vspace{-0.2in}
  \caption{Fidelity is plotted as a function of the atomic Rabi frequency $\Omega_{a}$ in GHz with all other parameters remaining same as in figure 3. }
  \label{fig 4}
\end{figure}

In Fig. \ref{fig 2} we have plotted the shifted and unperturbed phonon frequencies and the shifted equilibrium position as a function of atom-ion distance.
We consider that the ion-atom distance is $2.57\mu$m along the $x$-axis.
At this distance, the phonon frequency of the ion's x-axis is shifted to $2\pi\times 10.61$ MHz which is smaller than the harmonic frequency of unperturbed oscillation, which is  $2\pi \times 11.2$ MHz, as shown in Fig. \ref{fig 2}, but much larger than the linewidth of the ionic state. The unperturbed frequencies of the ion are $2\pi$(11.2, 18.2, 29.8) MHz, where frequencies in y and z direction are much higher than x-axis frequency \cite{Liebfried:RMP:2003}. This shift is sufficient to drive the system out of resonance with the first red sideband, thereby blocking the ion's motional transition.

We carry out our numerical work in the Lamb-Dicke regime, with $\eta=0.1$, such that the ion-phonon coupling remains in the perturbative regime while retaining sensitivity to shifts in trap frequency. We have set the Rabi frequencies for the transition $|0 \rangle \rightarrow |r \rangle$ for the atom is $\Omega_{a} = 2\pi \times 1$ GHz and the Rabi frequency for ion is $\Omega_{i} = 2\pi \times 1$ MHz leading to an effective sideband coupling of $\eta\Omega_{i}=2\pi \times 0.1$ MHz.
Detuning is chosen such that $\delta_{r} - \frac{1}{\hbar}V_{0} = 0$. At the distance $2.57 \mu$m, the ion-atom interaction is found to be $2\pi \times 0.2$ GHz. We choose such high Rabi frequency to avoid the excitation of higher phonon states which will ultimately lead to gate infidelity.

The dynamics of the system under the three pulse sequence is illustrated in Fig. \ref{fig 3} which shows the time evolution of the probability amplitudes of different states.  When the state is initially prepared in $|0,01\rangle$, it  evolves to $-i|r,01\rangle$ upon $\pi$ pulse on the atom. Now, as the atom is excited to $|r,01\rangle$ state, ion's phononic frequency changes. As a result, the transition between ionic states is blocked but $-i|r,01\rangle$ oscillates and acquires a phase when a $\pi$ pulse is applied on the ion. On application of a further $\pi$ pulse on the atom, the evolved state returns to its initial state provided the phase acquired previously is $\pi$. On the other hand, if  the system is prepared in $|1,01\rangle$ state, it is not influenced by the first $\pi$ pulse as state $|1\rangle$ is not coupled to $|r\rangle$. The state goes to $|1,10\rangle$ on application of second pulse as phonon frequency remains unchanged. Now when the third pulse is applied on the system it is not affected. As a result the final states transform into $|0,01\rangle$, $|0,10\rangle$, $-i|1,10\rangle$ and $-i|1,01\rangle$ from the initial states $|0,01\rangle$, $|0,10\rangle$, $|1,01\rangle$ and $|1,10\rangle$, respectively. Now,  we apply a {$\hat{S}$}-gate, a single qubit phase gate on atomic qubit.
We can thus realize CNOT gate. We calculate the gate fidelity using the formula given in \cite{Pedersen:PLA:2007} to be 89.94$\%$. Figure \ref{fig 4} shows the dependence of fidelity on the Rabi frequency of the atom. From this figure, it is worth noting  that the fidelity above 90$\%$ can be achieved provided $\Omega_{a}/2\pi > 1$ GHz. In general,  it is very difficult to increase Rabi frequency to GHz order. Nevertheless, the recent experiment by Huber {\it et. al.}  \cite{Huber:PRL:2011}  shows that Rabi frequency in the GHz regime is achievable in the case of Rydberg excitation.

\section{Conclusions and outlook}\label{Sec:4}

In this work, we have proposed and analyzed a hybrid atom-ion two-qubit quantum gate based on a Rydberg induced phonon blockade mechanism. By exploiting the long-range interaction between trapped ion and nearby Rydberg-excited neutral atom,  we have demonstrated that the ion's motional sideband transitions can be conditionally suppressed or enabled depending on the internal state of the atom. This work establishes that the state-dependent control of ion-phonon coupling provides a route to implement CNOT gate between atomic and ionic qubits.

The 90\% fidelity of our proposed ion-atom CNOT gate is much lower compared to the present-day state-of-the-art neutral atom and ion-trap CNOT gate fidelity which can exceed 99\%. However, as our numerical results indicate, higher fidelity  can be achieved if it is possible to increase the Rabi frequency of Rydberg coupling to a few GHz.
 Higher Rabi frequencies in the GHz regime may couple   nearby Rydberg sub-levels. For our chosen Rydberg S-state of $^{87}$Rb, the next nearby nondegenerate state is the D-excited state of the same principal quantum number ($n=90$). The energy difference between these two states is about 10 GHz as reported in reference \cite{Roy:OC:2022}. So with Rabi frequency being 1 GHz, and laser being tuned on resonance to the S Rydberg state, the probability of off-resonant coupling to the D-state is $0.5 \times \frac{1}{100}$ which is less than 1$\%$. So, the two level approximation will remain valid upto 1$\%$ error.  As far as generating strong ion-atom interaction through Rydberg excitation of the atom is concerned, this will lead to less than 1$\%$ error in the estimate of the interaction strength.
Clearly, increasing the Rabi frequency above 1 GHz will invalidate the two-level approximation. In that case, one possible remedy is to consider coupling to a pair of Rydberg levels by a single laser or a pair of lasers. In either case, our protocol requires an extension to include multi-level Rydberg excitation scheme while addressing the atomic qubit with laser pulses. This  will result in a superposition of the two Rydberg states. Then one has to take care of the third laser pulse to de-excite  back to the ground state from a superposition of a pair of excited Rydberg states. We hope to address these issues in our future communications. The bottom line is that the ion-Rydberg atom interaction should be sufficiently strong as to cause significant shift in the ionic oscillation frequency, yet remaining in the harmonic oscillation regime.

In our protocol we have not considered atomic center-of-mass motion (COM). Typically, the trapping frequency of optical tweezers is in the kHz order implying that the time period of atomic COM oscillations is in the  millisecond order. The size of optical tweezers is typically in $\mu$m order. With GHz order Rabi frequency,  Rydberg excitation will take place in nanosecond scale during which the atomic position uncertainty will be in pico-meter order which is negligible. However, the atom remains in the Rydberg state during the time when ionic qubit is addressed with the second laser pulse, with 0.1 MHz ionic Rabi frequency this time duration is of the order of 10 $\mu$sec. This means that the atomic position uncertainty during this time can be of the order of $10^{-2}$  $\mu$m. Compared to the atom-ion average separation of about 2 $\mu$m, this uncertainty will again lead to less than 1\% error in the estimation of ion-atom interaction.

In order to experimentally realize our proposed CNOT gate, one has to overcome several constraints. First placing an atom trapped in an optical tweezer at a precise distance from the center of the ion trap. Second, individually addressing both atom and ion qubits with appropriate optical pulses. Third, since our protocol requires atomic ground state-Rydberg Rabi coupling to be in GHz frequency regime, the question arises whether such high Rabi frequency can lead to other unwanted effects such as off-resonant coupling to other nearby Rydberg states. The first two constraints are related to technological developments - if right technology is available then any unwanted fluctuation in atom-ion distance can be minimized and also any imperfection in individual addressing can be overcome. The third constraint is technical and requires more careful system specifications.

We have presented a proof-of-principle theoretical demonstration of a hybrid ion-atom quantum computing protocol. Further extension and improvement of our protocol will open prospects for
applications of ion-atom hybrid quantum architecture in quantum technology.
A configuration of a linear chain of a few ions in a Paul trap and a few atoms in optical tweezers placed near each ion can serve as a node for distributed quantum computing. Different nodes can be connected by shuttling optical tweezers  and thereby transferring and distributing quantum information among different nodes.  In this context, it is worthwhile to note that  atomic and ionic qubits will respond to widely different laser frequencies. Another future direction would be to use such a distributed quantum computer as a node for quantum network by entangling the ionic qubits with photonic ones \cite{Canteri:PRL:2025}. In such a hybrid quantum network, one can use the best features of both charged and neutral qubits at one's disposal. For example, one can perform fast quantum gate operations in atomic qubits due to their scalability,  while ionic qubits may be used as quantum memory for their long coherence time and stability.

\bibliographystyle{unsrt}
\bibliography{bibliography}

\end{document}